\documentclass{article}
\usepackage[T1]{fontenc}
\usepackage{microtype,booktabs,array,amsmath,amssymb,graphicx,xcolor,xurl}
\usepackage[hypertexnames=false,hyperfootnotes=false]{hyperref}
\usepackage{algorithm,algorithmic,capt-of}
\usepackage[preprint]{mlsys2025}
\hypersetup{pdfsubject={Research preprint},pdfdisplaydoctitle=true}
\mlsyssetsymbol{equal}{*}
\mlsystitlerunning{PageWeaver: KV-Guided Query Unions for Sparse Attention}
\newcommand{\method}{\textsc{PageWeaver}}

\begin{document}
% Keep natural paragraph spacing on pages with large floats.
\raggedbottom
\twocolumn[{
\mlsystitle{PageWeaver: KV-Guided Query Unions for Sparse Attention}
\begin{mlsysauthorlist}
\mlsysauthor{Zhiyuan Li}{sensetime,equal}
\mlsysauthor{Zihan Li}{sensetime,equal}
\mlsysauthor{Zefang Yuan}{sensetime}
\mlsysauthor{Lei Wang}{sensetime}
\mlsysauthor{Hao Wang}{sensetime}
\end{mlsysauthorlist}
\mlsysaffiliation{sensetime}{SenseTime}
\mlsyskeywords{sparse attention, query grouping, KV reuse, FP8, GPU kernels}
\vskip 0.15in
\begin{abstract}
Dynamic sparse attention limits the KV pages selected by each query, but a small support does not necessarily yield efficient GPU work. Query unions share page loads and populate Tensor Core tiles; their cost depends on which queries are grouped together. We present \method{}, an execution design that uses selected-page affinity to assemble query groups while preserving each query's original support and complete output ownership. A bounded GPU search produces query IDs, and an ID-aware two-CTA kernel consumes them without materializing reordered Q tensors or cross-page partial outputs. A direct KV-page union implementation provides a complementary design study of nonlocal reuse and reduction cost. With FP8 KV throughout, the H200 Union8 implementation achieves a 1.70$\times$ geometric-mean complete-call speedup over the measured FlashInfer path on six captures. Online regrouping further lowers latency by 3.26--7.66\% on five selected 64K-context captures. Whole-model prefill throughput is 7.88--14.36\% above the tested native path; the incremental regrouping benefit is smaller, with observed median gains of 0.47--0.73\% at 32K/64K and regressions at 8K. A B300 comparison identifies cases where preparation cost and a stronger native kernel remove the advantage. These results separate execution-group reuse from the complete cost of exploiting it online.
\end{abstract}
\vskip 0.1in
}]
\printAffiliationsAndNotice{\mlsysEqualContribution\quad}

\begin{figure}[!t]
\centering\includegraphics[width=\columnwidth]{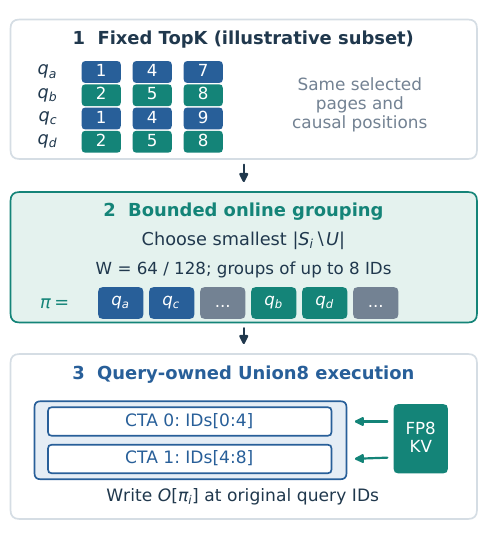}
\caption{\textbf{PageWeaver overview.} Fixed TopK pages guide bounded query regrouping. Original IDs carry support, causal positions, and output addresses through a two-CTA Union8 kernel. The illustrated page sets are a subset; measured groups use up to eight queries and Top16.}
\label{fig:overview}
\end{figure}

\section{Introduction}
Sparse attention selects a limited number of KV pages for each query, reducing the number of interactions relative to dense attention. Yet equal numbers of selected edges can produce very different execution costs. GPU Tensor Cores operate on matrix tiles, while dynamic page selection produces an irregular relation between queries and pages. Filling those tiles introduces a second scheduling problem: which queries should share a page traversal?

Query union is an established answer to part of this problem. Neighboring queries traverse the union of their selected pages, and membership masks retain each query's own support~\citep{litedsacode}. Grouping query heads supplies enough rows for a matrix tile and amortizes KV loads. Increasing the group size, however, can also enlarge the union and introduce more masked work. The choice of group size and the choice of group members are distinct optimization decisions.

\paragraph{Reuse must respect execution ownership.}
When adjacent queries share few pages, a page-oriented view can reveal reuse among distant queries. MSA's KV-outer design exploits such reuse by gathering queries for a page and merging partial attention outputs~\citep{msa}. We extend this view to pairs of pages sharing queries. Direct page union can reduce partial results, but also requires incidence inversion, work lists, and a final reduction. Our measurements show that this organization can win on controlled interleaved supports while losing on the examined model captures. Shared structure alone does not determine a profitable traversal.

\method{} instead uses the page relation to change the members of a query-owned group (Figure~\ref{fig:overview}). Within a bounded window, a GPU kernel greedily collects queries whose selected sets add few new pages. An ID-aware Union8 implementation then executes every query's complete attention and writes directly to its original output row. The selector is unchanged; membership and causal masks follow original query IDs. This design trades an online grouping cost for a smaller union without introducing cross-page output reduction.

\paragraph{Contributions and scope.}
We make three contributions. (1) We connect query-group size, page-set affinity, and output ownership in a common execution model, with direct KV-page union as a measured design alternative. (2) We implement bounded GPU regrouping and an ID-aware FP8 sparse-attention path on Hopper, retaining a complete query output within one CTA. (3) We separate base-kernel gains, incremental grouping gains, and whole-model prefill effects, including numerical comparisons and an auxiliary Blackwell study. The main evidence is an H200 MiniMax-M3 case study; selected operator captures and limited service repetitions constrain the generality of its conclusions.

\section{Related Works}
\paragraph{Selecting sparse support.}
MSA uses a block indexer and sparse main attention~\citep{msa}. BLASST prunes blocks using online softmax statistics~\citep{blasst}; LServe combines sparse attention with long-context serving~\citep{lserve}. These approaches choose or exploit a reduced set of interactions. Our input is an already selected support: we reorganize its execution without adding or removing any query's selected edges. BLASST's pruning rule and its quality calibration are therefore separate from the grouping problem studied here.

\paragraph{Grouping and sparse execution.}
LiteDSA unions adjacent queries' indices and applies per-query membership masks~\citep{litedsa,litedsacode}. MSA's KV-outer implementation chunks popular pages into bounded query work units and combines partial results~\citep{msa}. Both query grouping and page-oriented execution are existing techniques. Our direct page-pair path evaluates another point in this design space, while the final path changes query membership within a bounded search window and retains query-owned accumulation.

\paragraph{Reordering and kernel infrastructure.}
Sparse row reordering also seeks denser Tensor Core work. DTC-SpMM clusters rows by column-set similarity and cache reuse, with reordering amortized as preprocessing~\citep{dtcspmm}. Fused3S fuses sparse attention within row-owned blocks, schedules row windows by work count, and remaps operand layouts~\citep{fused3s}. Complete row ownership and similarity-based grouping are established techniques. Our focus is their bounded, per-call realization for dynamic page-selected attention: grouping is charged on every call, and original causal coordinates must survive the permutation. FlashAttention-3, FlashInfer, and FlexAttention provide complementary attention execution techniques~\citep{fa3,flashinfer,flexattention}. Our comparisons identify each backend's runtime and arithmetic rather than treating all FP8-KV kernels as interchangeable.

\section{Methodology}
\subsection{Fixed support, variable execution groups}
Consider one request and one KV head; its associated query heads share a selected-page set. Let $S_i$ be the valid, deduplicated logical pages selected by query $i$, $P$ the page size, $c_i$ its original causal position, and $L_i$ the valid context length. The supported tokens are
\begin{equation}
 A_i=\{t:\lfloor t/P\rfloor\in S_i,\;0\leq t<L_i,\;t\leq c_i\}.
\end{equation}
For each query head, with $z_{it}=q_i^\top k_t/\sqrt d$,
\begin{equation}
 o_i=\frac{\sum_{t\in A_i}\exp(z_{it})v_t}{\sum_{t\in A_i}\exp(z_{it})}.
\label{eq:attention}
\end{equation}
Head indices are suppressed. Invalid IDs contribute no edges, duplicate IDs do not duplicate probability mass, and empty support produces zero output. Logical page IDs are resolved through the request's physical KV map.

\begin{figure*}[t]
\centering\includegraphics[width=\textwidth]{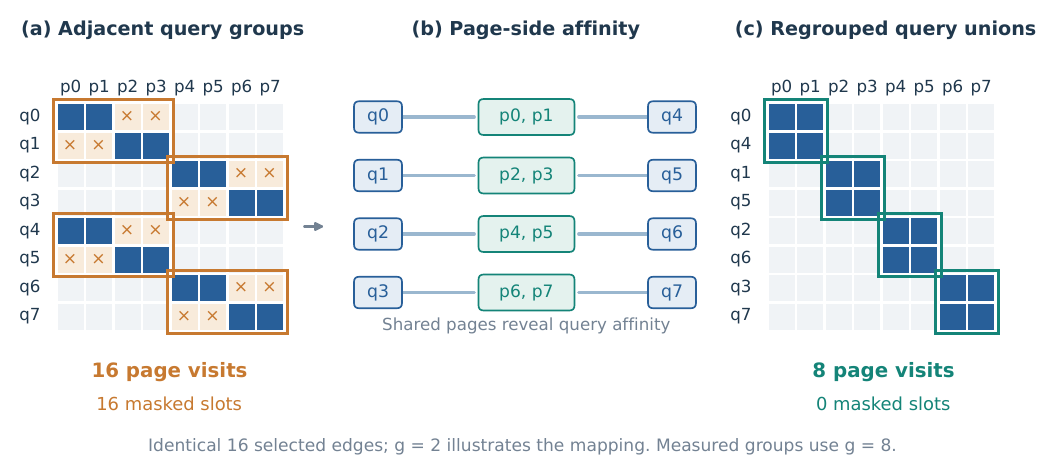}
\caption{\textbf{The same support can induce different amounts of work.} Adjacent groups mix two page communities; the page-side relation reveals nonadjacent queries sharing support. Regrouping reduces logical page visits from 16 to 8 and masked query--page slots from 16 to 0 in this illustrative example. All 16 selected edges are retained. The diagram uses two-query groups and two selected pages; the measured implementation uses groups of eight and Top16.}
\label{fig:incidence}
\end{figure*}

Define the incidence matrix $M_{ip}=\mathbf{1}[p\in S_i]$. For a group $G$ of $g$ queries, $U_G=\bigcup_{i\in G}S_i$ is its page union. Across a partition $\mathcal G$, logical page reuse is
\begin{equation}
 \rho_Q(\mathcal G)=\frac{\sum_i |S_i|}{\sum_{G\in\mathcal G}|U_G|}.
\label{eq:reuse}
\end{equation}
An execution rectangle has up to $g|U_G|$ query--page slots but only $\sum_{i\in G}|S_i|$ selected edges. Masks preserve support; masked Tensor Core slots are not generally free. Figure~\ref{fig:incidence} holds the selected edges fixed and illustrates the effect of changing the groups.

\subsection{Group size and output ownership}
Larger groups can improve reuse and tile utilization, but increase union expansion, staging, or synchronization. We compare implemented Union4 and Union8 configurations under matched arithmetic. H200 Union8 uses an eight-query transfer group and two four-query compute subgroups. Its logical transfer proxy is $\sum_G|U_G|$; a closer compute proxy is
\begin{equation}
 C_4=\sum_{G\in\mathcal G}\bigl(|U_{G_0}|+|U_{G_1}|\bigr),
\label{eq:subgroups}
\end{equation}
where $G_0,G_1$ are the two CTA subgroups. A subgroup can skip a page that none of its queries selects. These proxies describe different work: minimizing the full union need not minimize both subgroup rounds or balance the two CTAs. We use measured complete-call time to select the implemented group size.

\paragraph{The page-oriented alternative.}
Let $R_p=\{i:p\in S_i\}$. For a pair of pages $(a,b)$, $R_a\cap R_b$ identifies common queries, possibly far apart in sequence position. Accumulating both pages before writing a partial can save one partial per common query. For disjoint pairs, the potential reduction is $\sum_{(a,b)}|R_a\cap R_b|$. Exclusive queries still require single-page work, and every selected edge must be covered exactly once.

This organization partitions a query's output across page groups. If group $r$ produces maximum $m_r$, denominator $\ell_r$, and normalized partial $o_r$, let $\mathcal R=\{r:\ell_r>0\}$. For nonempty $\mathcal R$, the stable merge is
\begin{equation}
 o=\frac{\sum_{r\in\mathcal R} \ell_r e^{m_r-m}o_r}{\sum_{r\in\mathcal R} \ell_r e^{m_r-m}},\qquad
 m=\max_{r\in\mathcal R}m_r.
\label{eq:merge}
\end{equation}
An empty $\mathcal R$ produces zero output. Preparation, partial writes, and this reduction belong to the page path's cost. Our final design uses the same incidence information to find query neighbors while retaining complete query outputs in the query-oriented kernel.

\subsection{Bounded online grouping}
Page affinity between two queries is conceptually $(MM^\top)_{ij}=|S_i\cap S_j|$. We do not construct this dense matrix. Within a window of $W$ queries, Algorithm~\ref{alg:group} grows each group by minimizing the marginal union increase:
\begin{equation}
 i^*=\arg\min_{i\ \mathrm{unassigned}} |S_i\setminus U|.
\label{eq:greedy}
\end{equation}
Ties use original query order. The execution size is fixed at $g=8$, independently of the search window $W$. Windows remain within request boundaries and within the current prefill chunk. The method rebuilds IDs from current Top-$K$ sets on every call; it does not reuse page-pair decisions or previous requests' routes.

\begin{algorithm}[t]
\caption{Windowed page-affinity grouping}
\label{alg:group}
\begin{algorithmic}[1]
\REQUIRE Selected sets $S_i$, request boundaries, $W$, $g=8$
\ENSURE Request-local query permutation $\pi$
\FOR{each request-local window $B$}
  \IF{context exceeds the bitmap capacity}
    \STATE Append original query IDs of $B$ to $\pi$
  \ELSE
    \STATE $R\leftarrow B$
    \WHILE{$R\neq\varnothing$}
      \STATE $i\leftarrow\min R$; $G\leftarrow[i]$; $U\leftarrow S_i$
      \STATE $R\leftarrow R\setminus\{i\}$
      \WHILE{$|G|<g$ and $R\neq\varnothing$}
        \STATE $j\leftarrow\arg\min_{j\in R}(|S_j\setminus U|,j)$
        \STATE Append $j$ to $G$; $U\leftarrow U\cup S_j$
        \STATE $R\leftarrow R\setminus\{j\}$
      \ENDWHILE
      \STATE Append $G$ to $\pi$
    \ENDWHILE
  \ENDIF
\ENDFOR
\end{algorithmic}
\end{algorithm}

\paragraph{Preserving semantics.}
Execution slot $r$ reads $q_{\pi(r)}$, uses membership $M_{\pi(r),p}$ and causal position $c_{\pi(r)}$, and writes $o_{\pi(r)}$. The permutation is bijective within each request. Thus each original query sees exactly the same supported token set $A_i$ in Equation~\ref{eq:attention}; no probability mass or softmax state is shared between queries. This is a support-preservation argument in exact arithmetic. Quantization and reduction order remain separate numerical considerations.

\subsection{When grouping pays for itself}
Let $T_0$ be the complete original-order Union8 call, $T_{\rm fixed}$ the corresponding ID-aware call with a precomputed permutation, and $T_{\rm online}$ the call including GPU grouping. Define
\begin{equation}
\begin{aligned}
 S_{\rm exec}&=T_0-T_{\rm fixed},\\
 C_{\rm online}&=T_{\rm online}-T_{\rm fixed},\\
 S_{\rm net}&=S_{\rm exec}-C_{\rm online}.
\end{aligned}
\label{eq:net}
\end{equation}
The design is useful when $S_{\rm net}>0$. The difference $C_{\rm online}$ is an operational measure, not necessarily identical to a separately timed grouping kernel. Smaller unions may save traversal and arithmetic while nonadjacent IDs add address indirection or reduce locality. Bounded windows constrain both search cost and displacement. All required quantization, packing, metadata, attention, and output processing remain inside the measured call.

\section{Kernel Design}
\subsection{GPU grouping and ID-aware preparation}
One grouping CTA owns a window. W64 and W128 launch 64 and 128 threads; each query thread represents its Top16 pages with sixteen 32-bit bitmap words. Population counts of bitmap differences evaluate Equation~\ref{eq:greedy}. Warp reductions and a shared-memory winner reduction select the next query with deterministic tie breaking. The union resets after eight selections. Only an int32 query-ID array is produced; Q is not physically reordered.

The current capacity is 512 logical pages, or 65,536 tokens at the H200 page size of 128. Larger contexts use an identity permutation. Tail windows and partial groups preserve bijection. With $N$ queries and bitmap capacity $B$ pages, total grouping arithmetic scales as $O(NW\lceil B/32\rceil)$; independent windows run in parallel.

The union-preparation kernel gathers original Top-$K$ rows through these IDs, deduplicates group pages, and emits membership masks. Membership bits address execution slots; causal positions, request maps, and final stores address original queries. The existing FP8 KV packing path is retained and included in timing. No input-dependent CPU decision or host synchronization is added.

\begin{figure*}[t]
\centering\includegraphics[width=\textwidth]{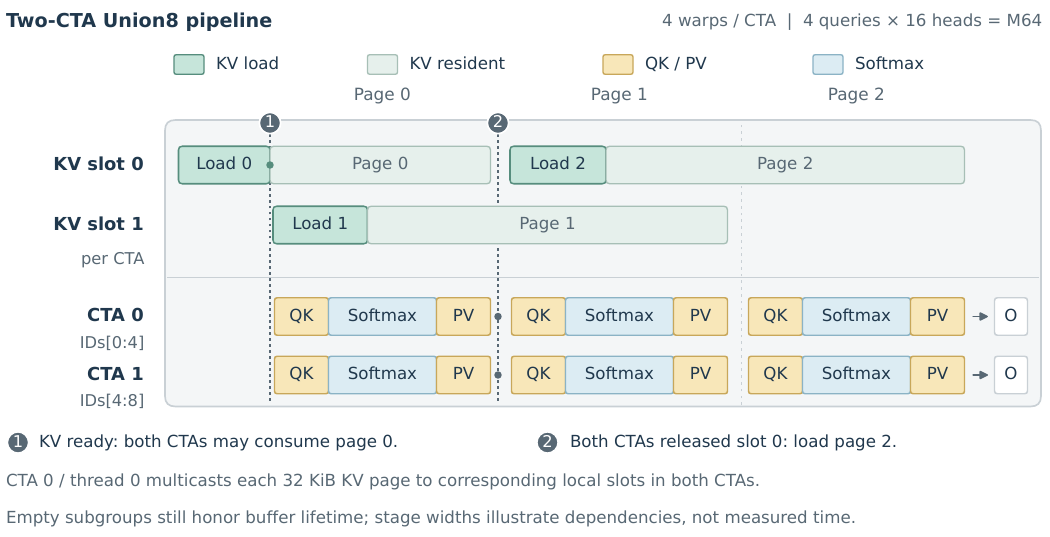}
\caption{\textbf{Hopper execution and buffer ownership.} Each CTA has four warps, four queries, and two local 32 KiB KV slots. Memory and compute lanes share an illustrative left-to-right schedule. Event 1 marks KV readiness; event 2 permits slot 0 to be refilled only after both CTAs release page 0. The same warps execute QK, softmax, and PV, retaining each query's state and original output ID. Empty subgroups still honor buffer lifetime. Widths are schematic, not measured durations.}
\label{fig:pipeline}
\end{figure*}

\subsection{Two-CTA Union8 execution}
The attention kernel launches clusters of two 128-thread CTAs (Figure~\ref{fig:pipeline}). Four queries times sixteen local Q heads provide 64 rows per CTA, mapping to an M64 Tensor Core tile. The cluster's eight-query union determines the page traversal. Each CTA applies its own four-query membership and causal masks, and retains FP32 running softmax statistics through the complete traversal.

Two KV staging slots separate readiness from reuse. The producer publishes staged data to the cluster; a slot becomes reusable only after the scheduled consumers finish. A CTA with no selected query on a page may skip its arithmetic but must still respect this lifetime protocol. Query grouping therefore changes the shared traversal and the number of useful subgroup rounds, rather than simply doubling independent CTA work.

The H200 path accepts BF16 Q and FP8 E4M3 KV. It quantizes Q per head-row for FP8 QK, keeps the softmax statistics in FP32, and scales probabilities before E4M3 conversion for FP8 PV. The compensating scale enters normalization; output is BF16. These arithmetic choices are fixed between original-order and regrouped Union8. Each CTA finishes its queries locally and stores directly to their original rows, eliminating a separate reorder-output pass and page-partial combine.

\subsection{Alternative execution and runtime boundaries}
The direct KV-page implementation inverts incidence, proposes disjoint page pairs, and forms common/exclusive query lists using full membership. Its CUDA producer accumulates paired pages into one local softmax state and writes a normalized partial plus log-sum-exp. A combine kernel finishes each query. This path evaluates a different execution organization; it changes work lists, producer mapping, and workspace together. Its selected padded partial workspace is approximately 1 GiB for 16K queries.

The SGLang bridge applies W128 regrouping to each current prefill chunk in all 57 sparse layers. Original request offsets remain authoritative in packed batches. The measured service has no learned or layer-selective scheduler. The B300 port maps the same IDs into a CuTe Union8 path with direct FP8 Q and physical page size 64. It includes the copy into a stable descriptor workspace required by that callable. These runtime and arithmetic differences are accounted for through within-platform comparisons.

\section{Experiments}
\subsection{Setup and evaluation questions}
We ask four questions: how much does the base Union8 execution improve over compatible kernels; how much additional work and time does online regrouping save; how much of the gain reaches whole-model prefill; and under what conditions does it disappear? H200 is the primary platform, with MiniMax-M3, FP8 E4M3 KV, Top16, head dimension 128, sixteen Q heads per local KV head, and page size 128. We use the fixed SGLang CUDA12.9 image and the pinned native SM90 implementation~\citep{sglq8}. FlashInfer is a compatible complete-operator baseline. The measured native interface is adapted to the fixed runtime without changing its device algorithm.

\begin{table}[t]
\centering\small
\caption{Evaluation units and their scope. Replays measure timing variability; they do not create additional independent documents.}
\label{tab:workloads}
\begin{tabular}{@{}>{\raggedright\arraybackslash}p{.25\columnwidth}>{\raggedright\arraybackslash}p{.68\columnwidth}@{}}
\toprule
Study & Workloads and repetitions\\
\midrule
Base execution & Six captures, one request; L3/L32/L59, 0/16K prefixes. Seven rounds $\times$ 20 replays.\\
Online grouping & Five captures of one constructed 64K input: L39/r0--3, L43/r1; plus one 16K L51 control. Eleven rounds $\times$ 50.\\
H200 service & Five frozen prompts per 8K/32K/64K bucket; concurrency 1/4; three measured blocks per arm/cell.\\
B300 auxiliary & Eight TP4/rank0 captures: L3/L20/L32/L59 at two chunks. Seven rotated graph/eager rounds.\\
\bottomrule
\end{tabular}
\end{table}

\paragraph{Measurement boundaries.}
Every complete operator call includes live preparation, quantization, packing, attention, and combine when required. CUDA Graph replays exclude compilation and reusable allocation; B300 also has eager event measurements. Backend order rotates on one device, with profiling collected separately. The service rotates three arms in one resident model after draining outstanding requests: native, original Union8, and online W128 plus Union8. It fixes TP8/DP2 with attention-TP4, effective 16K prefill chunks, disabled radix cache and CUDA Graphs, and one output token. There are 18 unmeasured warmup and 54 measured blocks, totaling 675 completed requests. Input IDs and token counts match across each cell's arms. Table~\ref{tab:workloads} distinguishes selected captures from independent workloads.

\subsection{Base execution: the value of a larger union group}
\begin{figure*}[t]
\centering\includegraphics[width=\textwidth]{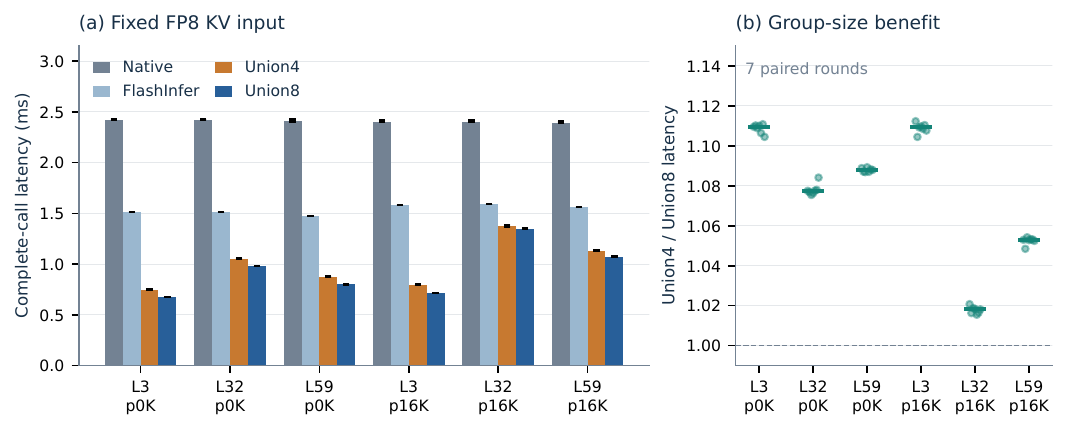}
\caption{\textbf{Base-kernel and group-size gains on H200.} Complete-call timings use the same six captures and FP8 KV bytes. Backend families can differ in Q/P arithmetic; Union4 and Union8 use matched arithmetic. Repeated-round ranges show measurement variability. L denotes layer and p the prefix preceding the current 16K-query chunk.}
\label{fig:base}
\end{figure*}

Figure~\ref{fig:base} separates total implementation performance from group-size effects. Across six captures, Union8 achieves a 1.701$\times$ geometric-mean speedup over FlashInfer and 2.662$\times$ over native. Although all receive FP8 KV bytes, backend families differ in Q and probability arithmetic. The matched-arithmetic Union4-to-Union8 comparison gives a 1.075$\times$ geometric-mean speedup, with bitwise-identical outputs on these inputs. Eight queries provide a useful Hopper execution unit; the result does not establish a universal optimum over group sizes.

\subsection{From nonlocal page reuse to query regrouping}
An eight-source development survey contains 2,736 correlated layer/rank/chunk observations across 57 sparse layers and four attention ranks. Per-source median reuse ranges from 3.85 to 4.54, with lower-overlap cases in particular later chunks. This motivates support-dependent grouping, but does not establish a universal layer policy or the frequency of performance wins.

\begin{figure*}[t]
\centering\includegraphics[width=\textwidth]{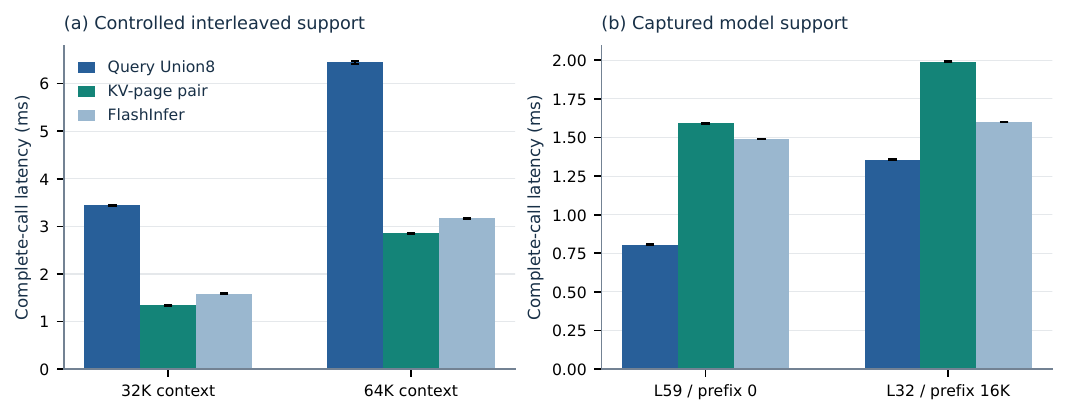}
\caption{\textbf{Direct page execution exposes an opportunity and its cost.} Controlled interleaved supports and model captures are distinct workload groups. Every timing includes required preparation and output processing. The constructed cases demonstrate a possible advantage of KV-page union; they do not estimate its prevalence in a model. Ranges cover repeated timing rounds.}
\label{fig:kv}
\end{figure*}

Controlled interleaved supports make the page-side opportunity explicit: adjacent queries select different page communities, while distant queries share them. Direct KV-page union lowers complete-call time relative to FlashInfer by 15.6\% at 32K context (1.340 versus 1.588 ms) and 10.1\% at 64K (2.852 versus 3.172 ms). Figure~\ref{fig:kv} contrasts these constructed inputs with two model captures, where Union8 is faster. In the deliberately low-overlap L43/r1 64K capture, page union also takes 1.827 ms versus 1.368 ms for Union8. Thus nonlocal reuse exists, but work-list and partial-output execution can absorb its value. This motivates using page relations for group membership while preserving query-owned accumulation.

\subsection{Online regrouping: saved work, paid overhead, net gain}
The five 64K captures have 15,272 current queries following a 49,152-token prefix, for 64,424 context tokens. Online W128 lowers complete-call latency by 3.26--7.66\% relative to original Union8; the 16K high-overlap control improves by 4.11\%. W64 and W128 grouping cost approximately 19.3--19.5 and 34.6--34.9 $\mu$s when timed alone. These costs are included in the online comparison. W64 wins over W128 on L39/r1, so a larger search window is not universally better.

\begin{figure*}[t]
\centering\includegraphics[width=\textwidth]{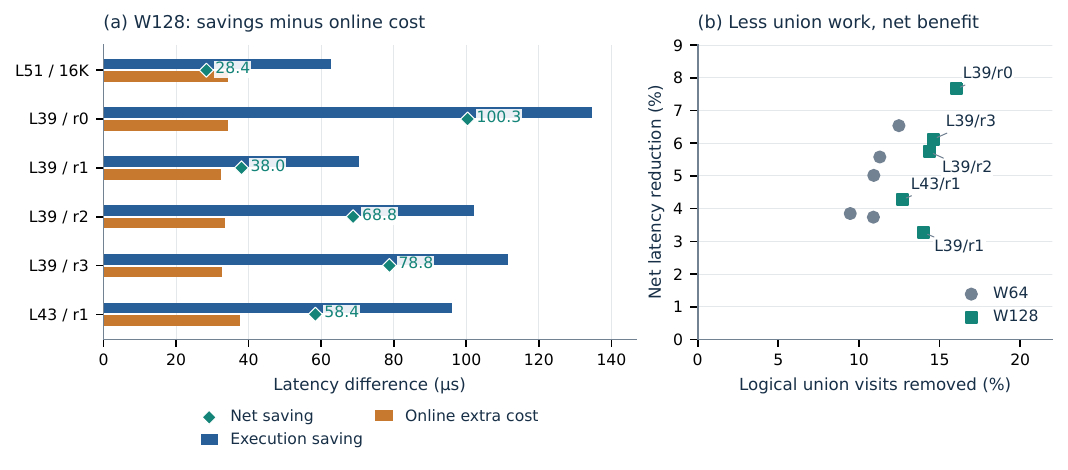}
\caption{\textbf{Explaining the incremental regrouping gain.} The complete online call is compared with original-order Union8 and execution using fixed grouped IDs on the same capture. Execution savings and the additional online cost determine the net saving (Equation~\ref{eq:net}). Structural statistics come from separate diagnostics of the same captures; page visits are logical counts, not HBM bytes. Timing repetitions do not expand the document sample size.}
\label{fig:mechanism}
\end{figure*}

Figure~\ref{fig:mechanism} exposes the cost balance rather than reporting only final speedups. On L43/r1, W128 reduces logical union visits from 102,429 to 89,426, or 12.70\%. Using medians from the same rotated experiment, original Union8 takes 1.36633 ms, execution with fixed grouped IDs takes 1.27037 ms, and online grouping plus execution takes 1.30792 ms. In Equation~\ref{eq:net}, execution saves 95.96 $\mu$s, the online increment costs 37.55 $\mu$s, and the net saving is 58.41 $\mu$s. The grouping-only measurement is 34.65 $\mu$s: it should not be substituted for the operational difference. Logical visits describe structural work, not measured HBM traffic.

\paragraph{Affinity and indirection matter.}
A separate L43/r1 diagnostic increases latency from 1.371 to 1.832 ms with random query IDs. Fixed W512 affinity IDs execute in 1.234 ms; materializing reordered Q/TopK and restoring output order takes 1.315 ms. This approximately 82 $\mu$s difference supports direct ID-based access. Global hash ordering followed by local grouping takes 1.298 ms, showing that similarity must be considered together with locality. W2048 reaches 1.185 ms execution time but requires 1.50 seconds of CPU search. These offline results measure execution potential, not deployable online speedup (Appendix~\ref{app:offline}).

\subsection{Whole-model prefill: total and incremental benefit}
\begin{figure*}[t]
\centering\includegraphics[width=\textwidth]{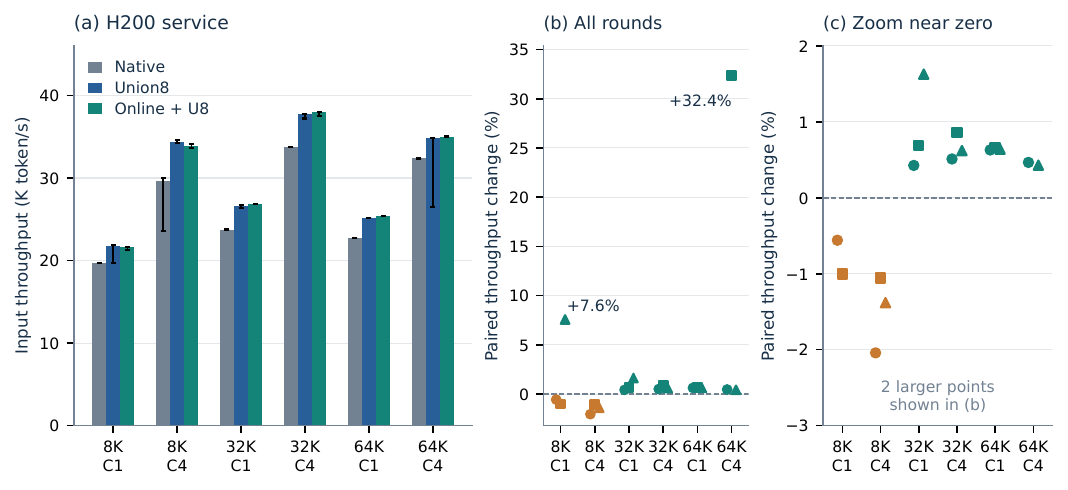}
\caption{\textbf{Whole-model H200 prefill and repeated regrouping effects.} Three arms use identical prompts within each length/concurrency cell. Throughput medians and ranges retain every block; the incremental view pairs online and original Union8 within each round. Large positive ratios arise when an original-Union8 block slows down, and remain visible. Three technical repeats do not establish statistical significance across documents.}
\label{fig:service}
\end{figure*}

Figure~\ref{fig:service} reports input-token throughput, including one generated token per request. The final path's median throughput is 7.88--14.36\% above the tested native service. Most of this advantage is already present in original Union8. Regrouping changes the median by +0.47--0.73\% at 32K/64K, and by $-0.56\%$/$-1.31\%$ at 8K with concurrency 1/4. These are observed median differences, not established population-level improvements.

Pairing the three 64K/C1 rounds yields +0.630\%, +0.666\%, and +0.641\%. Other cells show larger variation. For 64K/C4, one original-Union8 block drops to 26.51K token/s while the other two reach 34.82K and 34.83K; the paired ratios therefore include a +32.36\% outlier alongside approximately +0.43\% and +0.47\%. We retain it rather than filtering after observing the result.

Actual token ranges are 4,984--8,270, 30,563--32,842, and 63,997--65,666. Two of five unique prompts in the 64K bucket cross the bitmap capacity; affected chunks use identity IDs. All arms receive the same inputs, all requests complete without reported errors, and reported cached tokens are zero. No regrouping policy is selected per layer. The service comparison does not include FlashInfer; consequently it measures gain over the pinned native integration, not over every available serving backend. A complete service-stage trace and more independent repetitions are needed to attribute and validate the small increment.

\begin{figure*}[t]
\centering\includegraphics[width=\textwidth]{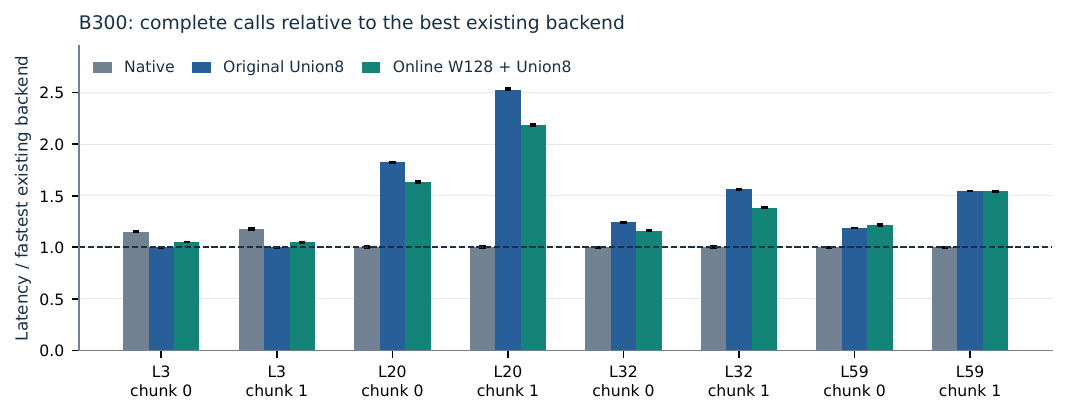}
\caption{\textbf{B300 auxiliary comparison.} Each input is normalized to its faster existing native/original-Union8 backend. Online grouping can improve Union8 while still losing to native. These are complete eager operator timings, not service throughput or a controlled comparison of H200 and B300 hardware.}
\label{fig:blackwell}
\end{figure*}

\subsection{Numerical checks and quality scope}
The implementation checks request-local bijection, membership preservation, causal and partial pages, invalid/duplicate IDs, empty support, packed requests, and fresh-input graph replay. GPU grouping IDs match the CPU greedy reference. Compute Sanitizer reports no memory errors. All six online H200 captures are bitwise identical to original Union8; this observation is limited to the tested inputs.

For output $O$ and reference $R$, we report $\mathrm{NRMS}=\|O-R\|_F/\max(\|R\|_F,\epsilon)$. Table~\ref{tab:precision} compares the base backends against the same Triton reference. Union8's differences are close to FlashInfer's on this set, while native differs more. These are numerical comparisons to a reference implementation, not downstream accuracy scores. On the separate online captures, a flattened FP32 reference sampled at 33 query rows per input yields NRMS of approximately 2.66--4.51\%; regrouping matches its parent output, including those sampled differences.

\begin{table}[t]
\centering\small
\caption{Numerical differences on six base-kernel captures. All rows use the same Triton reference and full output. Ranges are across inputs, not confidence intervals.}
\label{tab:precision}
\begin{tabular}{@{}lr@{}}
\toprule
Backend & NRMS (\%)\\
\midrule
Union8 & 2.246--3.967\\
FlashInfer & 2.288--4.077\\
Native SM90 & 6.258--9.120\\
\bottomrule
\end{tabular}
\end{table}

Preserving the selected edges does not remove the parent's FP8 approximation. The latest one-token service workload validates execution and throughput, not long-context generation quality. A paired task-quality evaluation of the final policy remains necessary before a quality-qualified deployment claim.

\subsection{B300 and the limits of regrouping}
The auxiliary B300 experiment uses CUDA13/SM103, direct FP8 Q, physical page size 64, and eight TP4/rank0 captures. W128 gives a 1.068--1.156$\times$ eager speedup over original Union8 on L20/L32, but native remains faster there. For L20's second chunk, original Union8, W128, and native take 1.497, 1.295, and 0.592 ms. On L3's first chunk, their times are 0.475, 0.500, and 0.547 ms: original Union8 already wins, and regrouping adds cost. Figure~\ref{fig:blackwell} shows that neither online window beats the faster existing backend on any of the eight tested inputs. No B300 serving experiment follows from this result.

The B port preserves support and has at most $1.55\times10^{-5}$ full-output NRMS relative to its parent. Its lack of an overall win bounds the current implementation, not page-affinity grouping in general. Native algorithms, input sets, Q entry formats, page sizes, and descriptor costs differ between the two platforms; hardware alone cannot explain the contrast.

\paragraph{Limitations and next tests.}
The operator evidence emphasizes selected inputs; independent natural-text validation should precede a general scheduling rule. The fixed 64K bitmap limits active regrouping on longer contexts. The present greedy objective minimizes the eight-query union, not the subgroup compute proxy in Equation~\ref{eq:subgroups}. A future selector could consider both, but no such cost-aware layer policy is evaluated here. More service repetitions, explicit stage attribution, a strongest-compatible service baseline, and final generation quality are the next evidence requirements.

\section{Conclusion}
\method{} uses the relation between selected queries and KV pages to improve execution groups while retaining each query's full attention output. A bounded online search and direct query-ID access connect this relation to a two-CTA FP8 Union8 kernel. The H200 study separates substantial base-kernel gains from a smaller additional regrouping benefit: the latter reduces complete-operator latency on selected inputs, but yields only modest and workload-dependent service changes. Direct KV-page execution and the B300 comparison identify why reuse must be evaluated together with preparation and output costs. The central design principle is to optimize the cost of executing a fixed sparse relation, including the cost of constructing its execution groups.

\bibliographystyle{mlsys2025}
\bibliography{references}

\clearpage
\appendix
\twocolumn[{
\begin{minipage}{\textwidth}
\section{Offline grouping diagnostics}
\label{app:offline}
\centering\includegraphics[width=\textwidth]{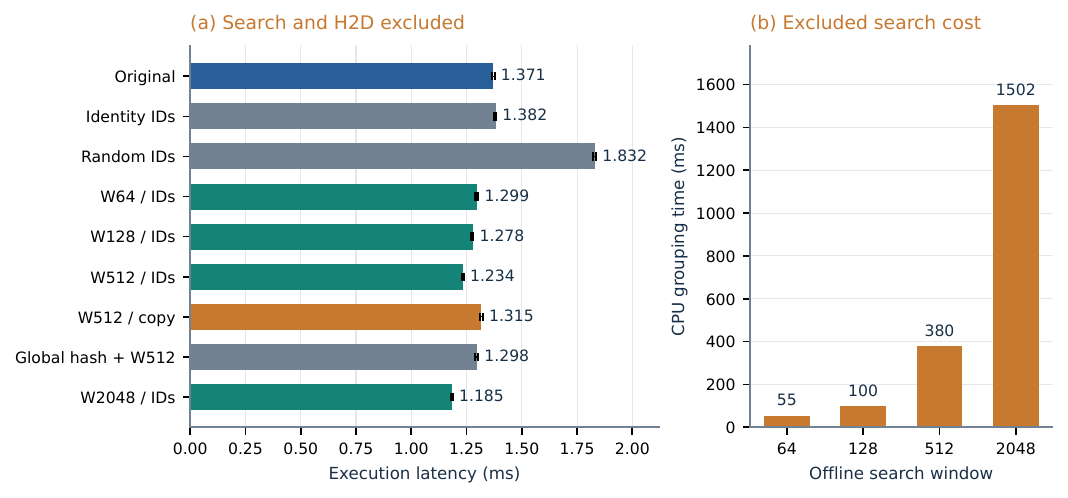}
\captionof{figure}{\textbf{Offline execution potential; grouping search excluded.} Fixed-ID and materialized variants use the same L43/r1 capture. Random IDs degrade locality and grouping; affinity reduces execution time. Materialization adds tensor movement. W2048's approximately 1.50-second CPU search is outside its execution bar and makes it unsuitable as an online implementation.}
\label{fig:offline}
\end{minipage}
\vskip 0.15in
}]

Figure~\ref{fig:offline} examines how affinity and data movement affect execution with fixed permutations on L43/r1. The CPU grouping search is excluded from every offline bar. Larger windows expose more potential reuse but can be prohibitively expensive to search. The W2048 CPU search takes approximately 1.50 seconds, compared with 1.185 ms for the resulting attention execution. It is an explanatory diagnostic rather than an online candidate.

\section{Reproducibility and interpretation}
The artifact contains the figure builders, source hashes, selected-kernel snapshots, per-round timing JSON, service block summaries, and numerical checks. Raw model weights and large captured GPU tensors are stored separately from the repository. Figures are regenerated from archived measurements; no new GPU run was performed to construct this manuscript revision.

\paragraph{H200 runtime.}
The image is the fixed SGLang \texttt{perf-m3-h200-d5658f11-cu129} build with digest \texttt{a1344259\ldots85631}; the artifact records the full digest. Native SM90 source is pinned to PR~37236 head \texttt{62bae610\ldots0347}. Source-hash verification and per-layer backend-hit logs check that service arms use their intended paths. Original and regrouped H200 Union8 both use two KV staging slots; the historical directory name \texttt{union8\_lb3} refers to launch bounds, not three stages.

\paragraph{Timing and sampling.}
H200 base-kernel results use seven rotated rounds of twenty graph replays; online grouping uses eleven rounds of fifty. Service cells use three measured blocks per arm, each containing $5C$ requests for concurrency $C$, drawn from five unique prompts. Median throughput is computed per arm and cell. Paired round ratios answer a different question and are shown separately. Neither replays nor repeated prompts create independent documents. Profiling is separate from performance timing.

\paragraph{Direct KV-page diagnosis.}
An earlier N256 Triton producer used 255 registers per thread, with 12.4\% achieved occupancy and 96.5\% L2 hit rate in one NCU run. These measurements motivated the later CUDA producer, but they are not a profile of that producer. The direct-page path changes several execution choices simultaneously; its results cannot isolate output ownership as a single causal variable. The main-text controlled workloads demonstrate a possible reuse pattern, while captured workloads retain their own provenance and comparison scope.

\end{document}